# Short Video Uprising: How #BlackLivesMatter Content on TikTok Challenges the Protest Paradigm

Yanru Jiang,[1] Xin Jin,[2] Qinghao Deng[3]

University of California Los Angeles, Department of Communication[1]
City University of Hong Kong,[2] Hong Kong Baptist University[3]
yanrujiang@g.ucla.edu, xin.jin@my.cityu.edu.hk, 19456301@life.hkbu.edu.hk

## Abstract

This study uses data from TikTok ($N = 8,173$) to examine how short-form video platforms challenge the protest paradigm established by the mainstream media in the Black Lives Matter movement, which was triggered by George Floyd's death on 25 May 2020. A computer-mediated visual analysis, computer vision, is employed to identify the presence of four visual frames of protest (riot, confrontation, spectacle, and debate) in multimedia content. Results of descriptive statistics and the t-test indicate that the three delegitimizing frames—riot, confrontation, and spectacle—are rarely found on TikTok, whereas the debate frame, that empowers marginalized communities, dominates the public sphere. However, although the three delegitimizing frames receive lower social media visibility, as measured by views, likes, shares, followers, and durations, legitimizing elements, such as the debate frame, minority identities, and unofficial sources, are not generally favored by TikTok audiences. This study concludes that while short-form video platforms could potentially challenge the protest paradigm on the content creators' side, the audiences' preference as measured by social media visibility might still be moderately associated with the protest paradigm.

## Introduction

On 25 May 2020, George Floyd, a Black American, was questioned by the police after being suspected of using a counterfeit $20 bill in a supermarket in Minnesota. A police officer forced Floyd to the ground and pressed his knees around Floyd's neck for seven minutes until Floyd stopped moving. He ignored Floyd's hopeless pleas for help and pushed a pedestrian, who tried to step forward to stop the violence, away (Selleck, 2020). Floyd's last words, "I can't breathe", have been heard throughout the world, igniting the American people's anger against racism and triggering another Black Lives Matter (BLM) movement in the country.

The BLM movement was initiated to increase attention from both the public and policymakers and advocates for Black Americans' rights and equal opportunities (De Choudhury et al., 2016). However, the mainstream media consistently fail to recognize protesters' contribution to promoting diverse voices and a more progressive and inclusive society (McLeod & Detenber, 1999). Mainstream journalists tend to portray protesters as violent and disruptive: they sensationalize social unrest, deeply setting the agenda as to how audiences perceive protesters and marginalized communities (McLeod & Detenber, 1999). This routine journalistic practice is known as the protest paradigm (Chan & Lee, 1984).

The proliferation of social media has enabled previously marginalized communities to speak out and share the backstory of their activism, an act that could fundamentally challenge the protest paradigm established by mainstream media (Harlow & Johnson, 2011). Studies on the role social media plays in the protest paradigm have primarily investigated more popular platforms such as Facebook and Twitter; however, these two platforms should not be the exclusive focus (Harlow et al., 2017). In fact, according to the Pew Research Center (2018), young people aged 13 to 25 years old are increasingly moving away from these mainstream platforms and predominantly use Instagram (and, particularly, Instagram stories), YouTube, Snapchat, and TikTok instead. The common features of these platforms are their multimedia capabilities, promptness, and facilitation of content produced by grassroots creators (Guinaudeau et al., 2020). Each of these features allows these platforms to be more effective in empowering youth and minorities, as well as in overthrowing the protest paradigm (Literat & Kligler-Vilenchik, 2019). During the summer of the BLM movement, the #blacklivesmatter hashtag suddenly became a trending topic on TikTok, receiving more than 4.9 billion views (Janfaza, 2020). Recognizing TikTok's worldwide popularity and its unique short-video multimedia format, this study focuses on the role of multimedia platforms in the BLM movement and looks specifically at how TikTok has avoided the stigmatization of protesters, enabled the context of the movement to be shared, and ultimately weakened the protest paradigm.

Considering the large amount of data labeling inherent in manual coding, we have adopted a computer-mediated visual analysis method, computer vision, for our study of TikTok videos to illustrate an efficient means of visual content analysis. Overall, this study sheds light on how short-form video platforms consistently challenge the protest paradigm on the content creators' side, especially since the capability of mainstream media like Facebook and Twitter remains controversial.

## Literature Review

### Protest paradigm in mainstream media

The Kerner Commission Report (1968), published 55 years ago, contends that the traditional media at the time was "shockingly backwards" in its coverage of race relations and protests, arguing that the media represented communities of color "through White men's eyes" and with a "White perspective" (Crittenden, 2019). The report highlights the inadequate representation of racial minorities in the corporate news publishing process, while White readers remained unaware of the difficulties faced by Black communities. Though the commission "suggested that greater representation might ameliorate this problem and provide more equal coverage" (Walker, 2018), this underrepresentation continues to be an issue as news corporations fail to adapt to the changing demographic composition of the country (The Kerner Report, 1968; Walker, 2018; Crittenden, 2019).

Similarly, social movement groups depend on media coverage to communicate their demands and shape public and official perceptions. However, Boyle et al. (2005) find that social movement groups face both selection and description bias in newspaper representation. The more a group deviates from the status quo, the more negatively the mainstream media will portray them (Boyle et al., 2005). In a 2019 study, Amenta et al. observe that the civil rights movement of the 1960s suffered compounded political legitimacy deficits in news coverage because the movement's leaders were not elected through political processes and held divergent views from the main political parties, substantively challenging the status quo. Fifty years after the publication of the 1968 Kerner Commission Report, racial justice activists still struggle to attain the fair representation and legitimization of marginalized groups in selective legacy news media coverage.

Previous research on media effects demonstrates a pattern in mainstream news practices in which news coverage sensationalizes social unrest and thus delegitimizes, or even demonizes, protesters' political participation in social movements (Stamps & Mastro, 2020). This pattern is known as the protest paradigm (Chan & Lee, 1984). McLeod and Detenber (1999) concludes that this coverage phenomenon in the mainstream media particularly serves the needs of the general audience by reinforcing its dominant perspectives. This dominant perspective has a tendency to preserve the status quo of the society while developing negative perceptions toward any groups that try to challenge the status quo (McLeod & Detenber, 1999). News reporters frequently quote from government officials, experts, elites, politicians or other mainstream media outlets to enhance their production efficiency and maintain their image of objectivity (Harlow & Johnson, 2011; Stamps & Mastro, 2020). However, authority figures are usually the very targets that protesters are confronting. There is a conflict of interest in directly quoting from these official sources (McLeod, 2007). Overall, news coverage tends to blame the individual players—the protesters—during demonstrations instead of questioning the systematic problems, such as discrimination, lack of equality and underpayment, that force disadvantaged groups to protest (Boyle et al., 2006).

McLeod and Detenber (1999) study framing to understand how the protest paradigm encourages support for the status quo among the dominant audience. Framing is the process of selecting some aspects of a story to make them more salient and imply a particular perspective for analyzing and interpreting the story (Entman, 1993). In the case of the protest paradigm, journalistic practices overemphasize the confrontation between protesters and police while ignoring the institutional players, such as policymakers, that protesters are targeting (Boyle et al., 2006).

Most social movements are organized by marginalized communities to draw the attention of authorities and the public, as well as to speak out on unfair treatment and lack of opportunities in society (Stamps & Mastro, 2020). Due to the protest paradigm, protesters from disadvantaged and vulnerable groups are further marginalized because the mainstream media fails to legitimize their participation and encourage diversity of views, ignoring the protesters' role in raising awareness and promoting equality in democratic society (McLeod, 2007; Weaver & Scacco, 2013). However, the ultimate goal of protest is to raise the awareness of the public and policymakers and to change existing policy that disadvantages marginalized communities. Advocates face a dilemma: if their political participation is mild, they probably will not even receive media coverage or have opportunities to voice their opinions (McLeod, 2007). This trend pressures protesters to adopt a relatively radical strategy and unavoidably face the risk of the protest paradigm (McLeod, 2007).

The double bind faced by marginalized communities is confirmed by Boyle et al. (2006) and McLeod's (2007) argument that the more radical the group is, the closely the news coverage follows the protest paradigm. Therefore, McLeod (2007) advocates for journalists to be trained to challenge the status quo of the protest paradigm in mainstream media. Besides journalistic development, Downing (2001) observes that alternative media, such as citizen journalism, activist publications and social media, might also challenge the protest paradigm established by the mainstream media.

## Protest paradigm on social media

While the protest paradigm in mainstream media has been extensively researched, with relatively consistent conclusions, its presence and degree of severity on social media is still controversial among previous studies (Harlow et al., 2020; Harlow & Johnson, 2011). There is a debate between two camps of scholars regarding whether social media has avoided the protest paradigm or, through accelerated communication, expanded its influence (Harlow et al., 2017; Ismail et al., 2019; Poell, 2014).

Social media, as an alternative to mainstream news, offers marginalized groups opportunities to challenge the suppression and unfair treatment they received from official and mainstream sources (Jenkins & Wolfgang, 2017). The platforms empower protesters with mobilizing resources to circumvent the censorship and gatekeeping that result from journalistic norms, and they ensure a diversity of voices online that overcomes barriers of time and space (Downing, 2001; Eltantawy & Wiest, 2011; Khamis & Vaughn, 2011). Ismail et al. (2019) propose the possibility that social media could even shift control of the conversation in the public sphere from elite-driven sources to grassroots actors. Additionally, Harlow et al. (2017; 2020) illustrate that social media users might prefer the legitimized portrayal of protests over the demonization and delegitimization frames. Overall, the authors observe that practices on social media diverge from the protest paradigm commonly found in the mainstream media.

However, Poell (2014) questions the effectiveness of social media by arguing that even protesters and grassroots content creators themselves may not avoid using mainstream protest frames for self-representation. In his research, Poell (2014) finds that protesters also frequently use spectacle and feature the crowds and dramatic scenes of protests on social media. Ismail et al. (2019) propose a different angle, noting that once a platform becomes dominant in the public discourse, the prominence of controversial, rebellious and niche narratives might be diminished. Additionally, multiple studies find that journalists, media outlets, elites and politicians dominate discussions of contested issues on Twitter (Barnard, 2012; Ismail et al., 2019; Wallsten, 2015). When grassroots groups and mainstream media coexist on social media, the latter might still have stronger influence in the public discourse over social movements (Paterson, 2013).

Accordingly, scholars have found differing evidence of the protest paradigm on various social media platforms. In their study, Harlow and Johnson (2011) conclude that a journalist's Twitter feed adheres less to the protest paradigm by instead legitimizing protesters and producing commentary and analytical content to explain the context of social movements. Moreover, Lasorsa, Lewis, and Holton's (2012) study suggests that, on Twitter, journalists from the mainstream media (such as national newspapers, television news, and cable news networks) are less likely to share user-generated content (UGC) and more likely to conform to the protest paradigm than those from the newcomer online-only sites. Harlow et al. (2017; 2020) also observe that Twitter audiences prefer more legitimizing content than what journalists have offered on the platform.

Comparing the news coverage and sharing of posts related to domestic and foreign protests on Twitter and Facebook, Kilgo et al. (2018) find that posts concerning domestic protests are shared more often than those regarding foreign protests on Facebook, while no significant difference is found on Twitter. Particularly, Facebook users tend to share more legitimizing information about protests that are farther away than domestic protests. These findings suggest that social media sharing and news coverage follow the same pattern of emphasizing the more delegitimizing elements of domestic protests and more legitimizing aspects of foreign protests.

When it comes to multimedia platforms, the content generated on video-sharing platforms such as YouTube and Vimeo is observed to have more legitimizing frames and feature peaceful protests more often than mainstream media outlets (Harlow et al., 2017). These platforms afford protest actors the ability to publish and share their own videos, empowering them to take control of the messages of social movements through their own UGC (Harlow et al., 2017).

Despite the emerging research on journalistic practice and news coverage on social media, a limited number of studies have directly examined whether the content produced on social media, rather than the online news circulated on these platforms, follows the protest paradigm. Due to the 140-character limit on Twitter, it is challenging for content producers to fully elaborate their perspectives and narratives. Harlow and Johnson (2011) conclude that although the Twitter account that they examine is the least likely of the sample sources to use delegitimization frames, it also substantially lacks sympathy frames, legitimizing frames, or injustice frames, which would favor protesters. Since the nature of social media might afford different norms of content production online, the cross-platform comparison of the presence of delegitimization frames requires more considerations.

**Grassroots Activism on TikTok**

Different platforms encourage different types of political expression and media exposure through their unique affordances (Maia & Rezende, 2016; Shifman, 2013). Although we are skeptical that mainstream social media like Facebook and Twitter have truly shed the protest paradigm, a cutting-edge short-form video platform like TikTok, which is popularly used by youth between 10 and 29 years old (who represent 47.4% of active TikTok users in the U.S.; Dean, 2022) and has high visibility for ethnic identities, could encourage the political expression of youth and marginalized groups (Literat & Kligler-Vilenchik, 2019). Differing from earlier multimedia platforms like Instagram and YouTube, which emphasize creators building a coherent online identity and brand, TikTok prioritizes virality and trendiness in individual videos and algorithmically pushes these videos to the front page, regardless of their creators' popularity (Abidin, 2020). This democratization of the attention economy for grassroots users and its encouragement of young and minority groups has led to TikTok's progressive visual content that truly avoids the protest paradigm. *This study aims at understanding whether the UGC on TikTok preserves the protest paradigm by mirroring traditional journalistic practices or if it challenges this paradigm by presenting an alternative grassroots-led source of information for protests and marginalized communities.*

In their research, Guinaudeau et al. (2020) illustrate multiple affordances that affect political expression and democratize the conversation on TikTok, including its televisual medium, the blurry line between content creator and audience, and the bottom-up trend of video-content production (Literat & Kligler-Vilenchik, 2019). As a short-form video platform, TikTok can disseminate frontline, real-time messages and appeal to emotions in ways textual information cannot (Harlow et al., 2017), enriching the diversity of content created by grassroots communities and protesters. One major difference between TikTok and other social media is that users see content in flows instead of stocks, which are the common layout of other platforms (Guinaudeau et al., 2020). TikTok does not directly hide users' profile information, but it increases profile anonymity through the flow layout. When video feeds are provided in a viral way on a scrolling feed, the source of these videos becomes less important than what content has been shown and recommended for users. Maia and Rezende (2016) claim that anonymity would affect online political expression. Specifically, anonymity encourages flaming, which the study authors contrast with civil or courteous discussion—an elite-driven and exclusive type of political discourse (Orbach, 2012)—resulting in the blurry line between elite-driven and grassroots-driven conversation on TikTok.

Furthermore, traditional media produces curated content based on journalists' selection and organizations' preferences, while multimedia platforms, especially TikTok, tend to feature unprocessed footage that reflects the different viewpoints of grassroots actors. In traditional mainstream media, the boundary between the content creator and audience is definitive and significant (Banjac & Hanusch, 2020); media outlets oversee content production and intervene in publication through gatekeeping in each stage to ensure the representation of their stance (Newman, 2009). This trend is still maintained among online versions of traditional mainstream media (Welbers et al., 2018), though media outlets are observed to gradually adapt to audiences' preferences (Lowrey & Woo, 2010). Consequently, content on mainstream media is usually well-structured, organized, and packaged (Newman, 2009).

On the contrary, social media redefine the transformation from audience member to content creator and blurs the distinction between these actors (Örnebring, 2008). Social media provide users with accessible technological tools, enabling ordinary users to create, upload, and publish content anytime and anywhere with low cost and barriers, especially for video-based platforms (Shah & Zimmermann, 2017). For example, the CapCut, an easy-to-use video editing tool introduced on TikTok (Liao, 2021), enhances users' video-making and publishing experience. When TikTok was first released, it encouraged users to produce videos up to 15 seconds with rhythmic and memorable pop song clips for strong virality (Matsakis, 2019). Later, the platform extended the video length limit to 60 seconds to promote more storytelling videos across a variety of topics (Matsakis, 2019). This UGC model contributes to the production of more raw footage, witness-style, unpolished videos that deliver grassroots individuals' perspectives. With the grassroots-led conversation and visual content format, protesters could reclaim the narratives of social movements through bottom-up, video-based communication aimed at educating the public on the reasons for their protest (Jones & Mattiacci, 2019).

Sharma (2013) finds that racialized communities, especially Black Americans, use social media for advocating against social injustice much more frequently than white Americans. Additionally, TikTok could further enhance Black Americans' identities through visual content (Serrano et al., 2020). Based on the grassroots-led

conversation and presence of Black American identities on TikTok, we propose the following hypotheses:

H1: TikTok will have a high proportion of videos created by unofficial accounts and a low proportion created by official accounts

H2: TikTok will have a high proportion of videos featuring Black American identities.

## Visual framing of protests

Most previous research on protest framing follows McLeod and Hertog's (1999) framework, which contains four dominant types of frames, including confrontation, riot, circus and debate. The frequent presence of the first three frames and lack of the debate frame in mainstream media confirms its use of the protest paradigm (McLeod, 2007). The *confrontation* frame focuses on the conflict between protesters and police. Harlow et al. (2017) describes the *riot* frame as the conflict between protesters and society. In the mainstream media, reporters overemphasize protesters' violent behaviors and simultaneously treat protesters as indistinguishable from rioters (Stamps & Mastro, 2020). Rantz (2020) argues that these two distinct forms of participation deserve differentiation and clarification. Protesters are aiming to raise awareness and draw more attention from stakeholders to reject the policy or structural issues that disadvantage the groups they represent (Castells, 2012); their peaceful forms of participation include rallying, marching, gathering, boycotting, forming human chains or patronizing shops (Stamps & Mastro, 2020). When participants become violent by assaulting and attacking innocents and police, destroying public property, and looting, newspapers should conceptualize them separately from protesters and identify them precisely as rioters (Stamps & Mastro, 2020). In addition to those two frames, the *circus* frame focuses on dramatic, violent and sensational scenes, such as crowds of protesters. Finally, the *debate* frame is encouraged by protest-paradigm researchers, who say that illustrating protesters' perspectives and addressing the backstory of the social movement could help the audience develop more empathy toward protesters and a more comprehensive understanding of the protest issues (McLeod 2007, Harlow et al., 2017).

In their analysis, Harlow et al. (2017) extend the application of protest framing to include online visual elements. Seeing the increasing trend in visual framing analysis, they design a method to identify the four protest frames in visual coverage. The riot frame is directly related to the violent behaviors during social unrest, and the salient elements of a riot include setting fires and destroying properties that can be easily identified from visuals. Accordingly, visual elements such as *protest* and *violence* could indicate rioting behavior. Confrontation can be identified with the presence of both *protesters* and *police*. Spectacle, which demonstrates the *large number of protesters*, is selected to represent the circus frame. In the debate frame, protest signs, as well as clips featuring voices of protesters and marginalized communities, are seen as a *dialogue* generated by protesters regarding their protest issues.

Harlow et al. (2017) observe that the spectacle and debate frames are more commonly found in the visual framing of protest multimedia content and coverage, across media outlets and countries on social media. This makes sense because users and media outlets grab the audience's attention by showing a large number of participants for a particular event (Harlow et al., 2017). In our investigation of TikTok visual frames, we expect to see a similar pattern of more debate frames, especially when TikTok enables a *duet* and mimic function that empowers politically active users to provide their own opinions, as well as react and respond to other TikTok videos, through user-generated content (Serrano et al., 2020).

Though TikTok presents various progressive features as a multimedia youth and grassroots centric platform, its impact on social movements or political discourse in general is still relatively underexamined. Therefore, we propose the following research questions:

RQ1: What will be the presence of the three delegitimizing frames (riot, confrontation, and spectacle) on TikTok?

RQ2: What will be the presence of the legitimizing frame (debate) on TikTok?

## Social media visibility and engagement

Guinaudeau et al. (2020) discuss four unique affordances of TikTok, including complete displaying of a visual feed, televisual medium, recommendation system and mobile-only user interface, leading to an extremely fast spread of information on this platform. At the same time, content creators can increase the visibility of their content by using hashtags, which allow audiences to find content based on their interests, further enhancing the spread of information on social media (Sharma, 2013). Liking, recommending and sharing have been fundamental elements of social media platforms, enabling users to like, share and comment on posts they have been exposed to (Harlow et al., 2017; Hermida et al., 2012). Usually, content that receives more views, likes, shares and comments has higher visibility based on the recommendation system, according to Gerlitz and Helmond (2011), though the mechanism for determining this high visibility varies across platforms. They observe that, even though every individual can voice a unique opinion on social media, only content that gains popularity can spread and reach a broader audience. Trilling et al. (2017) note that this system challenges the traditional newspapers, where journalists play gatekeeper roles, selecting which stories to distribute to the public. In contrast, on social media, the recommendation system becomes the new bottleneck for information flow, allowing only the most popular content to be seen by the public.

Beyond visibility, social media scholars further distinguish between different engagement levels and how they reflect different cognitive processes and motivations among users' views, likes, comments, and share engagements (Kim & Yang, 2017; Aldous et al., 2019). From lowest to highest levels of engagement, views or plays are a private behavior and likely to be determined by the recommendation system; likes indicates users' preferences, which result from the affective arousal of sensory and visual features in videos; comments and shares are usually associated with users' opinions and feelings (Kim & Yang, 2017; Aldous et al., 2019). Specifically, commenting is triggered by rational and interactive cognitive motivations, while sharing is motivated by affective or cognitive factors or a combination of both (Kim & Yang, 2017). From the activist point of view (Greenhow & Li, 2013), commenting and sharing behaviors on social media can facilitate collaboration and civic engagement on specific social issues within users' networks. Together, users can push a video to be featured by algorithms and shared with more users through likes, comments, and shares (Ghosh, 2021).

While previous studies support the idea that progressive multimedia platforms like TikTok could challenge the protest paradigm in content creators' perspectives (Harlow et al., 2017), not all creator accounts are equal in their ability to reach others and spread content, which means we do not have a consistent conclusion on whether social media audiences would adhere to the protest paradigm. In this study, we propose the following research questions:

RQ3: How will social media visibility and engagement, measured by views, likes, comments, shares, followers, and durations, vary regarding four protest visual frames (riot, confrontation, spectacle and debate)?

RQ4: How will social media visibility and engagement, measured by views, likes, comments, shares, followers, and durations, vary regarding sources of TikTok accounts and the presence of Black American identities?

## Method

### Data collection and processing

To ascertain the dominant frames of the BLM movement on TikTok, this study conducted a computer-mediated content analysis, employing computer vision. Generally, social media users tend to post opinions using corresponding hashtags to target the issues they focus on. Thus, this study used videos posted under relevant hashtags as the sample pool to conduct the data collection. According to TikTok's official *June Monthly Community Construction Report* (TikTok, 2020), #blm, #blacklivesmatter, #blackvoiceheard, #justiceforgeorgefloyd, and #protest were, at the time, the top five hashtags related to the BLM movement on TikTok. It should be noted that only around 2,000 videos can be stored on one hashtag page; therefore, only the top 2,000 videos, as measured by the number of views, were captured. Finally, the hashtag #protest included some irrelevant videos concerning, for example, protests in Turkey and Thailand, and these videos were hence removed from the study based on the text description of the video.

**Table 1.** *Distribution of videos and views of each hashtag*

| hashtags | videos | views |
|---|---|---|
| #blm | 1,868 | 18.7 billion |
| #blacklivesmatter | 1,887 | 22.8 billion |
| #blackvoicesheard | 1,948 | 685.4 million |
| #justiceforgeorgefloyd | 1,939 | 1.5 billion |
| #protest | 1,904 | 3.5 billion |

With the help of *Charles 4.5.6*, an HTTP monitoring software, a commonly used data collection tool for mobile applications, this study obtained a set of application program interface (API) URL addresses. Through these API addresses, the data dictionary for each video could be accessed and the video itself could also be downloaded and stored locally. After deduplication and limiting our time frame from 25 May 2020 to 15 October 2020, 8,317 unique videos were finally collected.

Although our data collection was limited to the top 2000 videos under each hashtag, our sample does not suffer from substantial selection bias. Our data represents a similar content composition, primarily comprising trending videos that are shown to general audiences. While the sample represents limited bias in terms of the audience, it may reflect content composition bias in terms of the content creators, such as by underrepresenting the content produced by grassroots producers.

### Computer vision modeling

For the computer vision processing element, this study first adopted a common practice by cutting each video into a set of JPEG images at one-second intervals, and, in so doing, essentially applied computer vision at the image level. Second, object detection and classification were conducted on these images to identify the visual elements. Since all TikTok videos have the same frame size, there was no need to resize the JEPG images, which may have led to a decrease in the object recognition reliability. Lastly, we derived the variables of each video by organizing the set of visual elements and ensuring that threshold parameters and the continuity of these parameters could be held.

**Measurement and operationalization**

This study used a computer vision approach to identify four visual frames of protest in each video, namely: riot, confrontation, spectacle, and debate frames. The riot frame usually includes protest violence and involves actions such as beating, smashing, looting, and burning. In this study, we used Won et al.'s (2017) dataset and protest violence detection model to identify violent scenes at the image level. The confrontation frame typically involves a situation where police appear at the protest and have conflicts with protesters. This study categorized videos with both police and protest scenes as containing confrontation frames. Python API *IdenProf* and Won et al.'s (2017) model were employed to identify police and protest scenes. The third type of frame, the spectacle frame, refers to images of a large crowd of protesters. This study used a Python API named *crowdcount* to detect the approximate number of people in each image. Finally, the debate frame relates to the demands and opinions of protesters and marginalized communities, and it involves sharing the backstory of social movements. The *face_recognition* API was used to detect the number of people and the percentage of area each person's head occupied in an image.

Python API's *deepface* and *face_recognition* were utilized to identify and count the number of Black Americans, respectively. If at least one Black American was detected in one image of a video, the video would be coded as containing a Black American "presence"; the video would be coded as containing a Black American "group presence" if there were more than two Black people.

TikTok accounts with a "verified" label are usually official accounts (i.e., belonging to an organization, official, government, or professional) and were thus coded as official sources. Accounts without this label are ordinary accounts registered by the general public and, as such, were coded as unofficial sources.

Play count (i.e., the number of views of a video), like count (i.e., the number of likes a video receives), comment count, share count, and duration can all be used to evaluate the social media visibility of a video.

**Content analysis**

Two coders with research backgrounds in social movement and communication studies conducted a content analysis on 500 videos, or around 10% of the data sample, to prepare both training and test datasets for computer vision models. The 500 videos were selected using stratified sampling, with the presence of each visual element exceeding 25% of the sampled videos.

The content analysis was conducted on four visual frames (riot, confrontation, spectacle, and debate) and one additional visual element (Black identity). Two coders coded 200 videos that were drawn from random sampling and passed intercoder reliability (ICR) tests based on the Cohen's kappa scores of the five visual elements: riot ($k = 0.78$), confrontation ($k = 0.74$), spectacle ($k = 0.79$), debate ($k = 0.85$), and Black identity ($k = 0.86$). Among all the ICR scores, the confrontation and debate scores were slightly lower than the others, but all five elements exhibited high levels of agreement between the two coders.

**Parameter fine-tuning**

Coding 500 TikTok videos based on the CodeBook, we selected 400 videos for fine-tuning parameters and thresholds of the five computer vision models and left 100 videos for validating the models. Based on the training dataset, we fine-tuned the parameters of the five models to arrive at the following accuracy scores of a balanced video sample: riot (73.75%), confrontation (77.50%), spectacle (76.75%), debate (73.50%), Black identity (71.50%), and overall accuracy (74.60%).

For riot frames, we fine-tuned Won et al.'s (2017) model by adjusting parameters and re-training 100 times for violence detection. If the violence score of more than three consecutive images (three seconds) exceeded 50%, then the video was coded as a riot frame. To constitute a confrontation frame, a video had to have four or more consecutive images exceeding 85% in terms of police presence and could not be detected as a debate frame. A video was coded as a spectacle frame if there were 150 or more people in three or more consecutive images. For debate frames, if there were fewer than five people in one video (determined by the maximum number of people detected in the video) and the largest person's head occupied more than 3% of the entire image area in six or more consecutive images, or if the largest person's head occupied more than 20% of the image area in at least three consecutive images, the video could be coded as presenting a debate frame. Since the model for detecting Black identities does not have additional parameters, it was omitted from this model fine-tuning process.

**Model validation**

By comparing the model labeling results to the ground truth of visual presence labeled by the coders, this study produced an accurate model for all five visual elements based on a balanced sample of 100 videos: riot (71.00%), confrontation (78.00%), spectacle (76.00%), debate (77.00%), Black identity (69.00%), and overall accuracy (74.20%).

## Results

According to results of descriptive statistics, 5.98% ($N = 489$) of videos were created by official accounts and 94.02% ($N = 7684$) of videos were created by unofficial accounts. H1 was supported. The proportion of videos featuring Black American identities (53.66%, $N = 4386$) was higher than videos without Black American identities (46.34%, *N* = 3787). H2 was also confirmed.

For four visual frames of protest, the proportion of videos with violence (7.93%, *N* = 648) was much lower than videos without violence (92.07%, *N* = 7525), which means the riot frame did not frequently appear in TikTok videos. There were very few videos in which police and protest appear simultaneously (0.80%, *N* = 65), which means there was almost no confrontation frame in TikTok videos. The spectacle frame illustrated a similar pattern: that is, videos containing crowd elements were very few in number (1.26%, *N* = 103), and videos without crowd elements (98.74%, *N* = 8,070) were dominant. The spectacle frame was also rarely found. Thus, RQ1 has been answered regarding the presence of riot, confrontation, and spectacle frames. In addressing RQ2 that is related to the debate frame, 45.38% (N = 3,709) of videos contained at least one person expressing his or her demands, opinions, or discussions regarding the BLM social movement, which was identified by a steady portrait on the screen. Additionally, Black American identities featured in 60.58% (N = 2247) of videos that contain a debate frame. H2 was further confirmed.

In addressing RQ3, a series of *t*-tests were conducted and results revealed that, videos without a riot frame had a significantly higher social media visibility in terms of play, $t(979) = -4.36, p < .001$, like, $t(1121) = -10.43, p < .001$, comment, $t(2357) = -6.81, p < .001$, and share counts, $t(2483) = -8.58, p < .001$ than videos with a riot frame. Videos without a confrontation frame presented a significantly higher social media visibility in terms of play, $t(77) = -4.77, p < .001$ and like counts, $t(72) = -6.09, p < .001$ compared to videos with a confrontation frame. The comparison of videos' comment and share counts were not significant. Videos without a spectacle frame had a higher significant social media visibility in term of the play count, $t(117) = -3.75, p < .001$. Videos with a debate frame, however, had a lower social media visibility in terms of play, $t(8077) = -8.23, p < .001$ and like counts, $t(8027) = -3.74, p < .001$, but the results was opposite for the share count, $t(7716) = 1.99, p < .05$.

Regarding RQ4, videos created and published by unverified accounts had a lower social media visibility measured by plays, $t(493) = 5.33, p < .001$; likes, $t(495) = 5.04, p < .001$; comments, $t(490) = 2.00, p < .05$. Videos with Black American identities had lower social media visibility in terms of the play count, $t(8171) = -2.93, p < .01$.

To further explore social media visibility in RQ3 and RQ4, we examined two visibility features from the content creators' side, including user types and duration of TikTok videos, with different visual frames. A series of *t*-tests indicated that videos without a riot frame and videos with a debate frame were created by accounts that have significantly more followers compared to videos with a riot frame, $t(1721) = -6.68, p < .001$, and videos without a debate frame, $t(6385) = 2.98, p < .001$, respectively. As for the length of content, videos without a riot frame or spectacle frame and videos with debate frames were significantly longer than videos with a riot frame, $t(814) = -7.65, p < .001$, videos with a spectacle frame, $t(105) = -2.31, p < .05$, and videos without a debate frame, $t(7300) = 25.23, p < .001$, respectively.

-- Insert Table 2 here –

From a more conceptual perspective, a broadly accepted criterion in the social media influencer marketing industry (Hawley, 2020) classifies users into three levels: ordinary users, mid-tier influencers, and celebrity influencers, based on their follower numbers. This user classification helps distinguish ordinary users from the so-called opinion leader. Regarding video durations, as we mentioned before, 15s and 60s durations are video-editing features of TikTok which generate two frequency peaks of video lengths in our sample (see Figure 1). These two peaks identified two video creation trends, short viral videos and long storytelling videos, that play different roles in visual framing.

**Figure 1.** *Frequency distribution of video length*

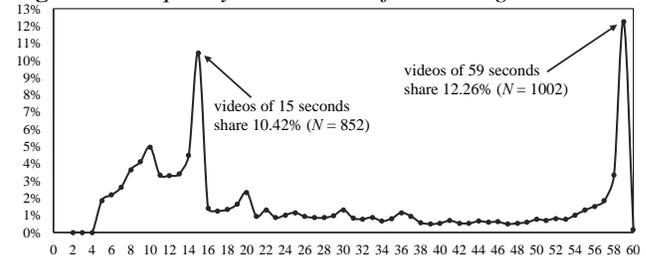

Based on the classification of user types and video types, we examined the chi-square distribution between these categories and the four types of visual frames (see Table 3). The results of the chi-square tests revealed consistent trends with *t*-tests. Particularly, mid-tier influencers are more likely than ordinary users to create videos with a debate frame, $c^2(2) = 194.3, p < .001$, videos without a riot frame, $c^2(2) = 98.62, p < .001$, and videos without a confrontation frame, $c^2(2) = 11.66, p < .01$. Videos longer than 45s are much more likely than videos shorter than 15s to present a debate frame, $c^2(2) = 529.56, p < .001$, and less likely to contain a riot frame, $c^2(2) = 49.49, p < .001$.

-- Insert Table 3 here --

## Discussion

This study examined TikTok's visual frames and the social media visibility and engagement of the BLM movement triggered by the death of George Floyd on May 25, 2020, through analyzing four common protest visual frames (riot, confrontation, spectacle and debate) identified previously. Overall, the frequency analysis, which summarizes the visual-frame composition on TikTok, proves our H1–H2 hypotheses and tentatively answers our RQ1 and RQ2 that progressive multimedia platforms like TikTok could potentially challenge the protest paradigm established by mainstream media. However, when the visibility of all protest frames is taken into account, the four indicators (views, likes, comments and shares) add another level of complexity to TikTok's effectiveness in challenging the protest paradigm. In general, the views, likes and comments indicators on TikTok still moderately adhere to the protest paradigm, while the shares indicator substantially avoids the protest paradigm and answers our RQ3 and RQ4.

Harlow et al. (2017) introduce four common protest visual frames—riot, confrontation, spectacle and debate— in their research on multimedia news shared online. The first three protest frames, which delegitimize protesters, are rarely found on TikTok under BLM-related hashtags. For the riot frame, only around 8% of all TikTok videos depict violent scenes. In general, TikTok users are not highly interested in portraying protesters as rioters or uploading videos that show the disruption of protests.

Confrontation, which describes the conflict between protesters and police, is only identified in fewer than 1% of videos. TikTok contains few videos that feature confrontations between protesters and police. Furthermore, some videos that feature police without protesters demonstrate a harmonious relationship between police and marginalized communities, contradicting the traditional confrontational dynamic between those two groups.

According to Harlow et al. (2017), the spectacle frame appears most frequently (64%) among all four types of frames in online multimedia news. Although the traditional protest paradigm established by mainstream news frequently uses the spectacle frame to sensationalize protests and demonize protesters, fewer than 1.5% of TikTok videos exhibit the spectacle frame. Since the TikTok platform allows users to construct bottom-up narratives for storytelling by featuring themselves, ordinary participants and bystanders (Guinaudeau et al., 2020), sensational scenes are less frequently captured in TikTok videos. Additionally, technical restrictions might contribute to the low proportion of spectacle frames. In news reporting, the common way to feature a crowd of protesters is with aerial photography and other professional photography methods. In comparison, TikTok is dominantly used by the general public who have limited ways to produce spectacle frames.

Over half of TikTok videos contain debate frames. Harlow et al. (2017) observe that the debate frame is the second most common frame (41.9%) among the four frames regarding multimedia news shared on social media. On TikTok, the debate frame substantially dominates the conversation particularly due to the ease of content creation. TikTok allows users to mimic trending videos, adopt fast-pace visual storytelling, and create *duet* videos as a reaction to others' content. All these unique features of TikTok enhance the possibility of creating a high-quality and entertaining debate video, ensuring the playfulness and sense of accomplishment of TikTok users. Additionally, almost 60% of debate videos include at least one Black American, which means debate frames are mainly created by the marginalized groups who advocate social change through social movements. These marginalized communities are more incentivized to tell the story and reasons for their protest, taking control of the protest narratives and educating the public on their long-term unfair treatment in society. Additionally, the hashtag #blackvoiceheard encourages Black Americans to feature themselves and describe the context of the BLM movement, substantially increasing the proportion of debate frames.

Similarly, more than half of all TikTok videos have at least one Black American, and around 20% of these videos contain more than one Black American. It makes sense that, in the BLM movement, more Black identities have been featured, and Black Americans themselves are actively participating in the TikTok discussion, taking control of the narratives and opposing the protest paradigm established by the mainstream media.

Only 6% of content is produced by verified TikTok accounts, and the rest, 94%, is generated by the general public. Mainstream media outlets have been criticized for favoring official sources of information for their news reporting, adhering to the protest paradigm. The algorithmic platform TikTok does not follow the pattern of favoring official sources. The blurry line between content creators and audiences, as well as bottom-up storytelling, on TikTok exposes the audience to more grassroots content. With the low presence of official sources to unofficial sources ratio, grassroots-led conversation can dominate the BLM discussion.

Content creators that emphasized the legitimizing elements of BLM protests and the debate frame, and downplayed delegitimizing aspects, such as riot and spectacle frames, tended to be the users with more followers and longer content. Distinct from Twitter, where elite opinion leaders such as journalists, media outlets, and politicians still dominate the conversation (Wallsten, 2015), the affordances and target users of TikTok suggest that its influencers tend to be younger and more grassroots-oriented. While TikTok celebrity influencers did not have distinguishable frame emphasis, the mid-tier influencers significantly featured more debate frame content and avoided riot and confrontation frames, promoting the legitimizing coverage of protesters to their followers through multimedia content. Concerning duration categories, short videos (videos shorter than 15 seconds) and long videos (videos between 46 and 60 seconds) have distinct natures, as TikTok guides users to select between 15-second and 60-second videos for sharing through its user interface design. While short videos tend to invite more sensational or surprising content, such as riot frames or a sequence of spectacle photos, long videos afford content creators sufficient time for political deliberation and storytelling, enhancing the presence of debate frames. Both a higher number of followers and longer videos from the content creator promote the visibility of legitimizing content of BLM on TikTok.

However, the series of *t*-tests performed on the social media engagement of protest frames, account sources, and Black American identities told a different story. Contradicting with the frequency analysis, videos with debate frames had less visibility as measured by views and likes compared to videos without debate frames. Despite the minority of videos originating from official accounts, they had higher visibility and longer durations. The audiences' preference for non-debate frames and videos created by officials might be a reflection of their inclination toward high-quality videos. Indeed, while debate frame videos cost less to produce, tend to be more informative, and can be created by any individual, they are disadvantaged by their need to compete for attention with other visual elements, such as violent and dramatics scenes.

On the positive side, videos with a debate frame were more likely to be shared by audiences and have longer durations, which was also true for videos including Black American identities (nevertheless, the shared variable of Black American identities was not significant). It is worth noting, furthermore, that these two types of videos tend to be longer because they allow their creators to explain the context of social movements through a debate frame. Black Americans substantially dominated the debate frame content by both producing and featuring in TikTok videos. Aligning with the frequency analysis, videos with a spectacle frame tend to have fewer views and shorter durations compared to videos without spectacle frames. The high visibility of non-spectacle videos indicated that, unlike journalists who usually illustrate dramatized scenes of protests, TikTok audiences did not find these elements of delegitimization interesting compared to other visual frames.

## Conclusion

Whether or not progressive multimedia platforms like TikTok challenge the traditional protest paradigm observed in mainstream media should be understood from the perspectives of two groups of users: content creators and audiences. From the content creators' viewpoint, TikTok presents a content composite of four protest-related visual frames that is fundamentally different from the content produced by mainstream media outlets, the online news shared on social media such as Facebook and Twitter, and video-sharing platforms (e.g., YouTube and Vimeo). The tremendous presence of debate frames and content produced by marginalized communities, such as Black Americans, elaborate on the contextual scene of the BLM movement. Additionally, extremely low proportions of riot, confrontation, and spectacle frames effectively avoid the delegitimization of protesters. When considering the interactive relationship between the four frames and user and video types, the legitimizing frame (i.e., debate frame) is closely associated with features that enhance content visibility, measured by follower numbers and video lengths. However, audiences' preferences, as suggested by their viewing, endorsing, commenting, and sharing behaviors, are not exactly consistent with content creators' preferences. Specifically, audiences' actions of viewing, endorsing, and commenting are, rather, moderately associated with the protest paradigm. When audience members conduct the highest level of user engagement and become content reproducers and content redistributors through sharing (Aldous et al., 2019), audiences are essentially able to legitimize protests and overthrow the protest paradigm.

The most important contributions—both theoretical and practical—of this research are as follows. Theoretically, this study fills the gap left by previous studies of the protest paradigm on social media: namely, that content produced by the general public, as opposed to news from official platforms circulated online, is rarely included in empirical studies due to the different content format of news and user-generated content. In our research, we emphasized four common visual frames related to protests and operationalized their identifications through computer vision tools. Moreover, while the effect of mainstream social media platforms like Facebook and Twitter on overthrowing the protest paradigm remains controversial, we used visual frames related to protests to investigate the multimedia platform TikTok, which consistently demonstrates its effectiveness in challenging the protest paradigm on the content creators' side. Practically, we used a cutting-edge computer vision method to detect the visual elements of social media videos and explored a set of threshold parameters and consecutiveness rules for automatic visual analysis. Finally, we observed a distinguishable preference between content creators and audiences based on our data analysis. It is proposed that future studies concerning the relationship between social media and protests should consider these two groups of users further.

Nonetheless, this study is limited in several aspects. First, due to the 2,000 quota on data collection of each hashtag, we only collected 8,173 videos in total. Furthermore, although the hashtags used in this study were recommended in TikTok's monthly report, it should be noted that not all videos related to the BLM movement may have been fully covered by the five hashtags we identified. Thus, our data may not fully reflect the entire conversation sparked by TikTok's content during the BLM movement. Because all the videos we sampled were "top 2000 trending videos" under each hashtag, we reserve our argument that the *t*-tests are generalizable to the population of BLM videos on TikTok. Another limitation is that since we used computer vision API for object detection and classification instead of developing our own training data, the accuracy of each image varied. However, we overcame this challenge by adopting a common practice of computer-mediated video analysis: identifying visual elements in consecutive images that pass thresholds and integrate the dummy variables to video-level information. By doing so, we improved the fault-tolerance rate considerably for video-level variables in our study. As this study used a facial recognition API for identifying racial identities, we acknowledge the harm and bias marginalized groups could suffer from the algorithm-driven identity detection including the African-American community. Future studies might want to utilize a multimodal approach, such as validating the identities based on descriptions of videos, to eliminate bias and unfairness that are inherent in algorithmic designs.

Moreover, although this research demonstrates TikTok's role in challenging the protest paradigm established in the mainstream media, it is worth exploring other multimedia platforms, such as Instagram (and particularly Instagram stories) and Snapchat, both of which are utilized by youth and Gen-Z users, in future studies. Lastly, protest signs are also recognized as a type of debate frame in social movements. While we did not address this visual element specifically in the study, future studies could discuss the representation of signs in both mainstream and social media visual content.

**Table 2.** Results of t-tests for protest frames, sources, Black American identities, and social media visibility and engagement

| | N (%) | Follower | | Duration | | Play | | Like | | Comment | | Share | |
|---|---|---|---|---|---|---|---|---|---|---|---|---|---|
| | | M (SD) | t (df) | M (SD) | t (df) | M (SD) | t (df) | M (SD) | t (df) | M (SD) | t (df) | M (SD) | t (df) |
| Riot | 648 (7.93%) | .17 (.67) | *** -6.68 (1721) | 23.33 (16.25) | *** -7.65 (814) | 1.06 (1.94) | *** -4.36 (979) | .16 (.29) | *** -10.43 (1121) | 2.83 (5.10) | *** -6.81 (2357) | 6.29 (14.23) | *** -8.58 (2483) |
| Non-riot | 7525 (92.07%) | .39 (1.86) | | 28.50 (19.39) | | 1.43 (3.19) | | .30 (.56) | | 4.76 (17.26) | | 13.15 (49.68) | |
| Confrontation | 65 (0.80%) | .26 (.97) | -.51 (8171) | 28.20 (18.21) | .05 8171 | .85 (.89) | *** -4.77 (77) | .14 (.20) | *** -6.09 (72) | 2.47 (2.89) | -1.04 (8171) | 5.25 (9.63) | -1.25 (8171) |
| Non-confront | 8108 (99.20%) | .38 (1.80) | | 28.09 19.22 | | 1.40 (3.12) | | .29 (.55) | | 4.62 (16.70) | | 12.67 (48.05) | |
| Spectacle | 103 (1.26%) | .50 (2.05) | .70 (8171) | 24.30 (16.75) | * -2.31 (105) | .89 (1.34) | *** -3.75 (117) | .22 (.31) | -1.30 (8171) | 3.03 (3.93) | -.97 (8171) | 10.80 (23.73) | -.39 (8171) |
| Non-spect | 8070 (98.74%) | .37 (1.80) | | 28.14 (19.23) | | 1.41 (3.13) | | .29 (.55) | | 4.62 (16.73) | | 12.63 (48.11) | |
| Debate | 3709 (45.38%) | .44 (2.13) | *** 2.98 (6385) | 33.83 (20.08) | *** 25.23 (7300) | 1.09 (2.96) | *** -8.23 (8077) | .26 (.53) | *** -3.74 (8027) | 4.58 (21.86) | -.14 (8171) | 13.77 (49.36) | * 1.99 (7716) |
| Non-debate | 4464 (54.62%) | .32 (1.47) | | 23.32 (17.04) | | 1.65 (3.20) | | .31 (.56) | | 4.63 (10.47) | | 11.64 (46.59) | |
| Verified | 489 (5.98%) | 3.78 (6.22) | *** 12.84 (488) | 29.89 (18.86) | * 2.14 8171 | 3.35 (8.58) | *** 5.33 (493) | .59 (1.31) | *** 5.40 (495) | 9.35 (55.78) | * 2.00 (490) | 13.10 (37.00) | .24 (8171) |
| Unverified | 7684 (94.02%) | .32 (.44) | | 27.98 (19.23) | | 1.28 (2.31) | | .27 (.45) | | 4.30 (9.76) | | 12.58 (48.49) | |
| Black | 4386 (53.66%) | .38 (1.46) | .40 (8171) | 31.74 (19.59) | *** 19.03 (8147) | 1.31 (2.74) | ** -2.93 (8171) | .29 (.55) | -.03 (8019) | 4.60 (11.27) | .01 (5569) | 12.74 (43.74) | .27 (8171) |
| Non-Black | 3787 (46.34%) | .37 (2.13) | | 23.86 (17.86) | | 1.51 (3.49) | | .29 (.54) | | 4.60 (21.22) | | 12.45 (52.26) | |

*p < .05, **p < .01, ***p < .001. follower (million), duration (second), play (million), like (million), comment (thousand), share (thousand).

**Table 3.** *Results of chi-square test for protest frames, user types, and video length*

|  |  | Riot | | Confrontation | | Spectacle | | Debate | |
|---|---|---|---|---|---|---|---|---|---|
|  |  | yes | non | yes | non | yes | non | yes | non |
| *User Type* | | | | | | | | | |
| ordinary user | actual | 486$_a$ | 4128$_b$ | 50$_a$ | 4564$_b$ | 62 | 4552 | 1784$_a$ | 2830$_b$ |
|  | expected | 366 | 4248 | 37 | 4577 | 58 | 4556 | 2094 | 2520 |
| mid-tier influencer | actual | 149$_a$ | 3160$_b$ | 13$_a$ | 3296$_b$ | 36 | 3273 | 1799$_a$ | 1510$_b$ |
|  | expected | 262 | 3047 | 26 | 3283 | 42 | 3267 | 1502 | 1807 |
| celebrity influencer | actual | 13 | 237 | 2 | 248 | 5 | 245 | 126 | 124 |
|  | expected | 20 | 230 | 2 | 248 | 3 | 247 | 114 | 137 |
| $\chi^2$ | | ***98.62 | | **11.66 | | 2.15 | | ***194.3 | |
| *Video Length* | | | | | | | | | |
| 1 ~ 15s | actual | 312$_a$ | 3315$_b$ | 25 | 3602 | 48 | 3579 | 1281$_a$ | 2346$_b$ |
|  | expected | 288 | 3339 | 29 | 3598 | 46 | 3581 | 1646 | 1981 |
| 16 ~ 45s | actual | 236$_a$ | 2127$_b$ | 25 | 2338 | 36 | 2327 | 990$_a$ | 1373$_b$ |
|  | expected | 187 | 2176 | 19 | 2344 | 30 | 2333 | 1072 | 1291 |
| 46 ~ 60s | actual | 100$_a$ | 2083$_b$ | 15 | 2168 | 19 | 2164 | 1438$_a$ | 745$_b$ |
|  | expected | 173 | 2010 | 17 | 2166 | 28 | 2156 | 991 | 1192 |
| $\chi^2$ | | ***49.49 | | 2.91 | | 4.10 | | ***529.56 | |

\*\**p* < .01, \*\*\**p* < .001. Ordinary user: accounts with less than 50,000 followers. Mid-tier influencer: 50,000~2,500,000 followers; Celebrity influencer: more than 2,500,000 followers. The degree of freedom for all eight (i.e., user type and video length × 4 visual frames) chi-square tests is the same ($df = 2$). The subscript letter a or b indicates a significantly unequal distribution at the .05 significance level (i.e., the actual value is significantly less or more than the expected value).